\documentclass[sigconf, nonacm, screen]{acmart}
\usepackage[turnoff]{notes}

\usepackage{hyperref}
\usepackage{boldline}
\usepackage{color}
\usepackage{float}
\usepackage{soul}
\usepackage[table]{colortbl}
\usepackage{tabularx}
\usepackage{bigstrut}
\usepackage[ruled]{algorithm2e}
\usepackage{amsmath,amsfonts}
\usepackage{amsmath}
\usepackage{booktabs}
\usepackage{graphicx} 
\usepackage{multirow}
\usepackage{array}
\usepackage{algorithmic}
\usepackage{balance}
\usepackage{blindtext}
\usepackage{caption}
\usepackage{hyperref}
\usepackage[noabbrev]{cleveref}
\usepackage{comment}
\usepackage{enumitem}
\usepackage{eqparbox}
\usepackage{fancybox}
\usepackage{fancyvrb}
\usepackage{framed}
\usepackage{graphicx}
\usepackage{ifthen}
\usepackage{listings}
\usepackage{mathrsfs}
\usepackage{mdwmath}
\usepackage{mdwtab}
\usepackage{multirow}
\usepackage{pifont}
\usepackage{textcomp}
\usepackage{url}
\usepackage{xspace}
\usepackage[normalem]{ulem}
\usepackage[framemethod=TikZ]{mdframed}
\usepackage{caption}
\usepackage{subcaption}
\usepackage{tcolorbox}
\tcbuselibrary{listings,skins}
\usepackage{mathtools}
\usepackage{blindtext}
\usepackage{lstautogobble}
\DeclareGraphicsExtensions{.pdf,.jpeg,.png}
\graphicspath{{figures/}}
\colorlet{lightgrey}{lightgray}

\newlength\Linewidth
\def\findlength{\setlength\Linewidth\linewidth
\addtolength\Linewidth{-4\fboxrule}
\addtolength\Linewidth{-3\fboxsep}
}
\newenvironment{examplebox}{\par\begingroup
  \setlength{\fboxsep}{3pt}\findlength
  \setbox0=\vbox\bgroup\noindent
  \hsize=0.90\linewidth
  \begin{minipage}{0.90\linewidth}\normalsize}
    {\end{minipage}\egroup
    \textcolor{gray20}{\fboxsep1.2pt\fbox
     {\fboxsep5pt\colorbox{gray05}{\normalcolor\box0}}}
    \endgroup\par\noindent
    \normalcolor\ignorespacesafterend}

\usetikzlibrary{shadows}
\usepackage{graphics}
\newmdenv[
    tikzsetting= {fill=blueish},
    skipabove=0.33em,
    skipbelow=0.33em,
    linewidth=1pt,
    innerleftmargin=4pt,
    innerrightmargin=4pt,
    innertopmargin=2pt,
    innerbottommargin=2pt,
    linecolor=gray95,
    roundcorner=2pt, 
    shadow=true,
    shadowsize=4pt,
    shadowcolor=gray95
]{questionbox}

\newmdenv[
    tikzsetting= {fill=greenish},
    skipabove=0.33em,
    skipbelow=0.33em,
    linewidth=1pt,
    innerleftmargin=4pt,
    innerrightmargin=4pt,
    innertopmargin=2pt,
    innerbottommargin=2pt,
    linecolor=gray95,
    roundcorner=2pt, 
    shadow=true,
    shadowsize=4pt,
    shadowcolor=gray95
]{answerbox}

\newmdenv[
    skipabove=0.33em,
    skipbelow=0.33em,
    innerleftmargin=4pt,
    innerrightmargin=4pt,
    innertopmargin=2pt,
    innerbottommargin=2pt,
]{lessonbox}

\usepackage{tikz}
\newenvironment{lesson}
{
    \begin{lessonbox}
}
{\end{lessonbox}}

\newenvironment{result}
{\begin{answerbox}}
{\end{answerbox}}

\newenvironment{question}
{\begin{questionbox}}
{\end{questionbox}}

\definecolor{javared}{rgb}{0.6,0,0} % for strings
\definecolor{javagreen}{rgb}{0.25,0.5,0.35} % comments
\definecolor{javapurple}{rgb}{0.5,0,0.35} % keywords
\definecolor{javadocblue}{rgb}{0.25,0.35,0.75} % javadoc

\lstdefinestyle{basejava}{
  language=java,
  showstringspaces=false,
  basicstyle=\small\ttfamily,
  keywordstyle=\bfseries\color{javapurple},
  commentstyle=\itshape\blue,
  identifierstyle=\blue,
  frame=none,
  backgroundcolor=\color{white},
}

\lstdefinestyle{CustomJava}{
  belowcaptionskip=\baselineskip,
  breaklines=true,
  xleftmargin=\parindent,
  language=java,
  showstringspaces=false,
  basicstyle=\scriptsize\ttfamily,
  keywordstyle=\bfseries\color{javapurple},
  commentstyle=\itshape\blue,
  identifierstyle=\blue,
  belowskip=1pt,
  numbers=left,
  gobble=0
}

\lstdefinestyle{CustomJavaWoNumbers}{
  belowcaptionskip=0.5\baselineskip,
  breaklines=true,
  xleftmargin=\parindent,
  language=java,
  showstringspaces=false,
  basicstyle=\scriptsize\ttfamily,
  keywordstyle=\bfseries\color{javapurple},
  commentstyle=\itshape\blue,
  identifierstyle=\blue,
  belowskip=0.5pt,
  numbers=none,
  gobble=0
}

\lstdefinestyle{codit}{
  belowcaptionskip=\baselineskip,
  breaklines=true,
  language=java,
  showstringspaces=false,
  basicstyle=\scriptsize\ttfamily,
  keywordstyle=\bfseries\color{javapurple},
  commentstyle=\itshape\blue,
  identifierstyle=\blue,
}

\newcommand\blue[1]{\textcolor[rgb]{0.00,0.00,1.00}{{#1}}}

\definecolor{blueish}{RGB}{250, 250, 255}
\definecolor{greenish}{RGB}{250, 255, 250}
\definecolor{redish}{RGB}{255, 200, 200}

\definecolor{highlight}{RGB}{175, 255, 100}
\definecolor{gray01}{gray}{.98}
\definecolor{gray05}{gray}{0.95}
\definecolor{gray08}{gray}{0.92}
\definecolor{gray10}{gray}{0.90}
\definecolor{gray12}{gray}{0.88}
\definecolor{gray15}{gray}{0.85}
\definecolor{gray18}{gray}{0.82}
\definecolor{gray20}{gray}{0.80}
\definecolor{gray25}{gray}{0.75}
\definecolor{gray30}{gray}{0.70}
\definecolor{gray35}{gray}{0.65}
\definecolor{gray40}{gray}{0.60}
\definecolor{gray45}{gray}{0.55}
\definecolor{gray50}{gray}{0.50}
\definecolor{gray55}{gray}{0.45}
\definecolor{gray60}{gray}{0.40}
\definecolor{gray65}{gray}{0.35}
\definecolor{gray70}{gray}{0.30}
\definecolor{gray75}{gray}{0.25}
\definecolor{gray80}{gray}{0.20}
\definecolor{gray85}{gray}{0.15}
\definecolor{gray90}{gray}{0.10}
\definecolor{gray95}{gray}{0.05}
\xspace%

\newcommand{\Comment}[1]{}

\newcommand{\linecode}[1]{\lstinline[escapechar=@,basicstyle=\ttfamily]{#1}~}

\newcommand{\tick}{\ding{51}}

\newtcbox{\inlinebox}[1][]{enhanced,
 box align=base,
 nobeforeafter,
 colback=blueish,
 size=small,
 left=0pt,
 right=0pt,
 boxsep=2pt,
 #1}

\newcommand{\lessons}[1]{
    \begin{lesson}
        \tick~#1
    \end{lesson}
}

\newcommand{\RQ}[2]{%
    \begin{question}
        \label{rq-#1}
        \noindent\textbf{{RQ{#1}.~#2}}
    \end{question}
}

\newcommand{\RS}[2]{%
    \begin{result}
        \textbf{\hyperref[rq-#1]{Result {#1}}:~}{#2}%
    \end{result}
}

\renewcommand{\cref}[1]{\Cref{#1}}
\newcommand{\ic}[1]
{\begin{small}\texttt{#1}\end{small}}
\newcommand{\completion}{\ic{code-davinci-002}\xspace}
\newcommand{\gpt}{\ic{gpt-3.5-turbo}\xspace}
\newcommand{\edit}{\ic{code-davinci-edit-001}\xspace}

\newcommand{\rom}[1]{\uppercase\expandafter{\romannumeral #1\relax}}

\newcommand{\etal}{\hbox{\emph{et al.}}\xspace}
\newcommand{\eg}{\hbox{\emph{e.g.,}}\xspace}
\newcommand{\ie}{\hbox{\emph{i.e.,}}\xspace}
\newcommand{\wrt}{\hbox{\emph{w.r.t.}}\xspace}

\newcounter{hypothesis}
\usepackage{listings}
\usepackage{color}

\definecolor{dkgreen}{rgb}{0,0.6,0}
\definecolor{gray}{rgb}{0.5,0.5,0.5}
\definecolor{mauve}{rgb}{0.58,0,0.82}

\newcommand{\rqa}{Can LLMs generate fixes from natural language intent for NL2Fix?} 
\newcommand{\rqb}{What kind of candidate fixes do LLMs generate?}
\newcommand{\rqc}{ What sources of information do LLMs need to generate fixes for NL2Fix?}
\newcommand{\rqd}{Can LLMs be used to rank fixes for NL2Fix?}

\newcommand{\nlTocode}{{\it nl2code}}
\newcommand{\nlToedit}{{\it nl2edit}}
\newcommand{\nlTofix}{{\it nl2fix}}
\newcommand{\dforjnltoedit}{{\it Defects4J-Nl2fix}}

\title{Towards Generating Functionally Correct Code Edits from Natural Language Issue Descriptions}

\begin{document}

\author{Sarah Fakhoury}\authornote{Equal contribution.} 
\affiliation{%
  \institution{Microsoft Research}
  \country{}
  }
  
\author{Saikat Chakraborty\footnotemark[1]{}}
\affiliation{%
  \institution{Microsoft Research}
  \country{}
  }
\author{Madan Musuvathi}
\affiliation{%
  \institution{Microsoft Research}
  \country{}
  }

\author{Shuvendu K. Lahiri}
\affiliation{%
  \institution{Microsoft Research}
  \country{}
  }

\begin{abstract}
    
Large language models (LLMs), such as OpenAI's Codex, have demonstrated their potential to generate code from natural language descriptions across a wide range of programming tasks. Several benchmarks have recently emerged to evaluate the ability of LLMs to generate functionally correct code from natural language intent with respect to a set of hidden test cases. This has enabled the research community to identify significant and reproducible advancements in LLM capabilities. However, there is currently a lack of benchmark datasets for assessing the ability of LLMs to generate functionally correct {\it code edits} based on natural language descriptions of intended changes. This paper aims to address this gap by motivating the problem \nlTofix{} of translating natural language descriptions of code changes (namely bug fixes described in Issue reports in repositories) into correct code fixes. To this end, we introduce \dforjnltoedit{},  a dataset of 283 Java programs from the popular Defects4J dataset augmented with high-level descriptions of bug fixes, and empirically evaluate the performance of several state-of-the-art LLMs for the this task. Results show that these LLMS together are capable of generating plausible fixes for 64.6\% of the bugs, and the best LLM-based technique can achieve up to 21.20\% top-1 and 35.68\% top-5 accuracy on this benchmark.

\end{abstract}

\maketitle

%%%%%%%%%%%%%%%%%%%%%%%%%%%%%%%%
\section{Introduction}
\label{sec:intro}

There has been a recent surge of interest in using large language models (LLMs) to accomplish a variety of software-development tasks. For instance, GitHub Copilot~\cite{copilot_2022}, powered by OpenAI's Codex model~\cite{codex_2021}, has demonstrated impressive capabilities in generating code based on a natural language description and related code contexts. Recent additions to Copilot~\cite{copilot_x_blog_3_28_2023} also highlight transforming code based on natural language instructions.
Besides Copilot, LLMs are powering several other commercial AI-assisted software development products such as Amazon CodeWhisperer~\cite{codewhisperer}, GhostWriter~\cite{ghostwriter} and Tabnine~\cite{tabnine}.

Research on LLMs for code generation has been spurred by benchmarks that are used to evaluate and often improve the performance of these models on particular tasks.
For example, the HumanEval~\cite{codex_2021} benchmark for evaluating the Codex model was instrumental in guiding the training of the much improved GPT-4 successor by OpenAI~\cite{openai2023gpt4}.
Improvement on such offline benchmarks often translates to improvement in real-world usage of LLMs for similar tasks (say, in an IDE).
In fact, for the task of code generation from natural language (we hereby refer to it as \nlTocode{}), several additional benchmarks have come up in recent years, including basic crowd-sourced coding benchmarks such as MBPP (from Google)~\cite{austin2021program}, Code Competition benchmarks~\cite{li2022automating}, and  those based on real-world programs~\cite{yu2023codereval}.

While several such benchmarks capture the ability of LLMs to generate functionally correct code from natural language intent, there is a lack of benchmarks that evaluate an LLMs ability to perform {\it edits} based on a natural language intent (we refer to as \nlToedit).

In real-world software development, a user is much more likely to  perform such software evolution and maintenance tasks in the context of an existing software repository compared to  writing a self-contained program from scratch using natural language. 
Besides, in modern software development with cloud-hosted continuous integration (CI) pipelines, mature software programs often evolve through {\it Pull Requests} (PR) made in response to a natural language {\it Issue} description corresponding to bug-fixes, optimizations, feature additions etc.

In this paper, we take the {\it first step} towards creating a benchmark for \nlToedit{} and evaluating current state-of-the-art LLMs on the problem.
In particular, we focus on the restricted problem of \nlTofix{} that consists of the task of fixing a buggy program where the bug is described in a natural language within an Issue description.
We leverage the popular Defects4J~\cite{just2014defects4j} benchmark comprising of real-world bugs and their fixes to define the \dforjnltoedit{} benchmark for \nlTofix{}.
We introduce an augmented dataset of 283 Java programs with their high-level descriptions of intended code changes, along with the suite of tests that ensure that the fix addresses the defect while preserving the functionality.
We also introduce metrics for evaluating LLMs for the problem of \nlTofix{} on this benchmark.

\begin{figure*}[t]
\includegraphics[width=0.88\linewidth]{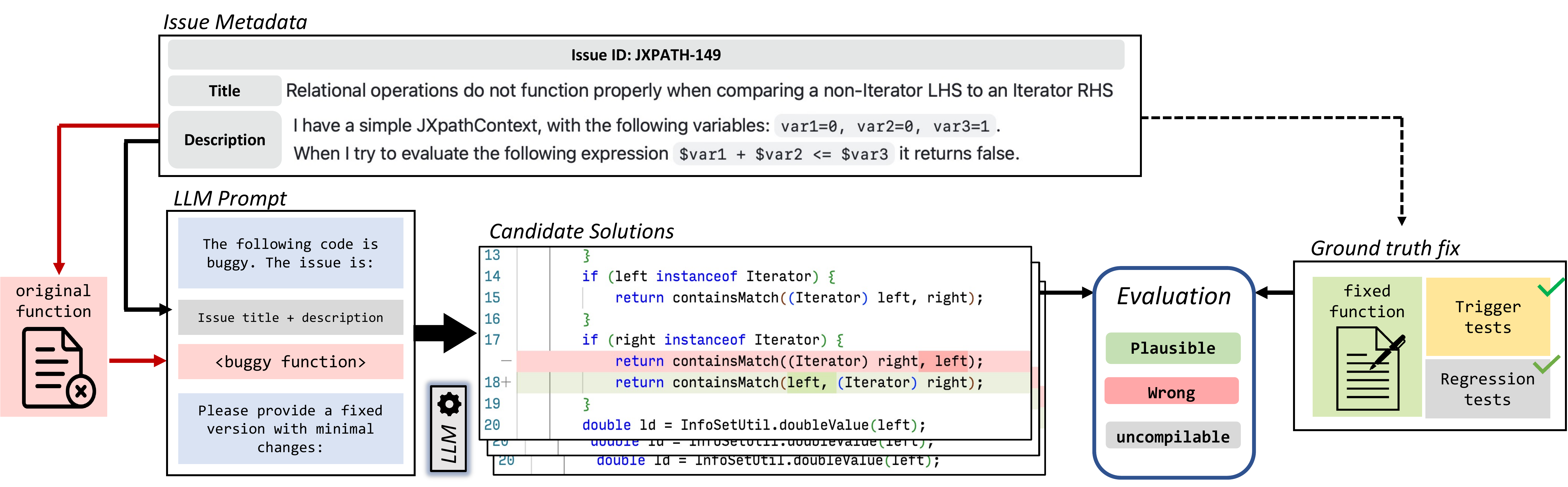}
    \caption{Overview of the NL2Fix problem setting. Illustrated is the issue title, description, and plausible fix for the JXPATH-149 bug.  The figure demonstrates the standard LLM prompt used to generate candidate patches, and the evaluation of the patches using ground truth trigger and regression tests. }
    \label{fig:overview}
\end{figure*}

Figure~\ref{fig:overview} illustrates an example from the \dforjnltoedit{} dataset. 
It shows an {\it Issue} containing the \texttt{Bug ID} that references an issue from the JxPath\footnote{https://issues.apache.org/jira//browse/JXPATH-149} project hosted on Jira\cite{jira}; a \texttt{Title} that describes the bug at a high level, and finally a \texttt{Description} that provides more details of the bug including a natural language description of a scenario to reproduce the bug. 
The \texttt{original function} consists of the method body that contains the fix --- we restrict our dataset to fixes that are localized to a single method.
The Issue, along with the  buggy function, constitute an input to the \nlTofix{} problem.
The output patches generated are evaluated for functional correctness and defect-freedom using the \texttt{Regression tests} and the \texttt{Trigger tests}, respectively.
Each test $t$ in the set of \texttt{Trigger tests} fails on the buggy program and passes the user-provided fix in the form of \texttt{fixed function}.

The motivation to use hidden tests only for validation comes from several directions.
First, prior work observes that most real-world bug-fix Issues are not accompanied by a failing or a trigger test~\cite{kang2022large}. Second, although a program may have accompanying regression tests, running such tests in CI pipelines can be expensive and can only be invoked a small number (say, 5) times to be practical.
As discussed later in related work (Section~\ref{sec:related}), keeping the tests hidden distinguishes the \nlTofix{} problem from the problem of automated program repair (APR)~\cite{apr_survey_monperrus_acm_2018,apr_survey_gazzola_2019,gao2022program}.

This paper also contributes a detailed empirical evaluation of the performance of current state-of-the-art (SOTA) LLMs on this dataset.
We choose three different flavors of LLMs based on the generative pre-trained transformers (GPT) neural architecture from OpenAI, (a) the Codex code completion model \ic{code-davinci-002}, (b) the Codex code editing model \ic{code-davinci} \ic{-edit-001}, and (c) the ChatGPT conversational model \ic{gpt-3.5} \ic{-turbo}.
We evaluate these models under different sampling settings and {\it prompting} strategies and perform detailed quantitative and qualitative analysis on the accuracy and quality of the suggested fixes.

Our results demonstrate that these LLMs together are capable of  generating {\it plausible} patches (\ie satisfy the regression and trigger tests) for a significant fraction, 64.6\% of the bugs when sampling for up to 100 candidates.
More interestingly, the ChatGPT model \ic{gpt-3.5-turbo} outperforms other models in terms of both the pass@1, pass@5, and pass@100 accuracy metrics (we describe the pass@k metric more formally later in Section~\ref{sec:passk}).
Finally, we describe a generic approach to rank an unordered set of candidate patches based on  LLM computed {\it embedding similarity}; the ranking makes the suggestions deterministic with top-1 and top-5 accuracy of 21.20\% and  35.68\%, respectively.  

These findings highlight both the non-trivial nature of the \dforjnltoedit{} benchmark, as well as the capabilities of current LLMs to form decent initial baselines that can spur further research.

\textbf{Contributions.} In summary, in this paper:
    ($i$) We motivate the \nlTofix{} problem and present a non-trivial augmented benchmark \dforjnltoedit{} along with metrics.
    
    ($ii$) We perform an extensive empirical evaluation of the performance of three state-of-the-art LLMs on this benchmark.
    
    ($iii$) We describe a ranking strategy based on embedding similarity to provide a ranked and deterministic list of fixes.

\section{Research Questions}
\label{sec:exp}
This paper aims to understand the performance of current state-of-the-art LLMs on the problem of \nlTofix{}: the task of fixing a buggy program from natural language intent. To this end, we define four research questions to empirically evaluate model capabilities:

\RQ{1}{\rqa}
To answer this RQ we explore the pass@k accuracy of three SOTA LLMs for generating correct bug fixes using natural language issue descriptions. Additionally, we extract quantitative statistics about the generated fixes, including: the prevalence of duplicate suggestions, compilation percentage, and the distribution and overlap of unique bugs for which there are plausible patches from each approach. 
\RQ{2}{\rqb}

To shed light on the nature of LLM generated candidate patches, we study the characteristics of the patches in the context of their similarity to the developer written ground truth fixes and the input buggy code. 
\RQ{3}{\rqc}

We explore what level of information LLMs need in order to correctly generate fixes from natural language descriptions. We experiment with different prompting styles using curated information, including: the high-level issue summary, in-depth issue description, 0-shot and 1-shot prompt settings, as well as bug fix reasoning generated using Reasoning Extraction prompting strategies. 

\RQ{4}{\rqd}

Based on observations from RQ1, RQ2, and RQ3 we explore how LLMs can be used to design a simple ranking approach to  rank fixes from the unordered set of candidate fixes, allowing better approximation of pass@k metrics needed for a developing a deterministic and practical bug fix recommender.

\section{Approach}
\label{sec:method}

\begin{table}[t]
\centering
\caption{Statistics of the Dataset}
\label{tab:dataset_statistics}
\resizebox{\linewidth}{!}%
{%
\begin{tabular}{lr|rr|rr|rr}
\toprule
\multirow{2}{*}{\textbf{Project}} & {\multirow{2}{*}{\textbf{\# Bugs}}} & {\textbf{SH$^\dagger$}} & {\textbf{Avg.}} & \multicolumn{2}{c|}{\textbf{Avg. Change}} & \multicolumn{2}{c}{\textbf{Issue   Length}} \\
 &  & {\textbf{Bugs}} & {\textbf{hunks}} & {\textbf{Line}} & {\textbf{Token}} & {\textbf{Title}} & {\textbf{Desc.}} \\
 \midrule
Chart & 6 & 5 & 1.17 & 1.83 & 9.67 & 7.17 & 149.33 \\
Cli & 28 & 16 & 1.68 & 4.07 & 22.07 & 9.36 & 206.46 \\
Codec & 11 & 9 & 1.27 & 2.18 & 10.82 & 12.73 & 171.09 \\
Collections & 1 & 1 & 1.00 & 1.00 & 1.00 & 20.00 & 457.00 \\
Compress & 36 & 19 & 1.78 & 5.39 & 29.28 & 10.03 & 320.42 \\
Csv & 12 & 8 & 1.42 & 2.50 & 18.92 & 10.83 & 104.25 \\
JacksonCore & 13 & 9 & 1.38 & 3.69 & 20.69 & 11.38 & 251.69 \\
JacksonDataBind & \multirow{1}{*}{67} & \multirow{1}{*}{36} & \multirow{1}{*}{1.87} & \multirow{1}{*}{5.37} & \multirow{1}{*}{33.90} & \multirow{1}{*}{11.90} & \multirow{1}{*}{294.49} \\
% Databind & & & & & & &\\
JacksonXml & 5 & 1 & 2.80 & 6.20 & 30.80 & 11.60 & 126.80 \\
JxPath & 10 & 5 & 1.60 & 4.80 & 22.60 & 9.70 & 159.10 \\
Math & 73 & 37 & 2.11 & 5.29 & 35.00 & 10.07 & 165.18 \\
Mockito & 21 & 16 & 1.33 & 3.29 & 27.90 & 9.00 & 311.67 \\
\midrule
\textbf{Overall} & 283 & 162 & 1.78 & 4.65 & 28.76 & 10.53 & 231.92\\
\bottomrule
\end{tabular}%
}

{\footnotesize $^\dagger$SH: Single-Hunk}
\end{table}

\subsection{Dataset}

In this paper, we take the {\it first step} towards creating a benchmark for a \nlToedit{} and evaluating current state-of-the-art LLMs on the problem. We focus on the restricted problem of \nlTofix{} which consists of the task of fixing a buggy program where the bug is described in natural language within an issue description. 

We choose the Defects4J~\cite{just2014defects4j} benchmark, comprising of bugs and tests from real-world issues, from which we can extract issue descriptions.

In particular, we use Defects4J 2.0,  a well-known benchmark of 835 manually curated real-world bugs and fixes gathered from 17 Java projects. The existing dataset consists of a set of bugs, bug reproducing test cases (trigger tests), and regression test cases which load the class in which the method under test is contained. Each bug in the Defects4J dataset contains a \ic{PRE\_FIX\_REVISION} and \ic{POST\_FIX\_REVISION} version that represents the buggy/fixed versions of the code respectively. The two versions reflect the actual state of the project when the bug was discovered/fixed.

We use these developer-written tests to evaluate generated patches, a patch must pass both the trigger and regression tests to be considered a plausible patch. While this does not guarantee semantic equivalence between a generated patch and the ground truth fix, we argue that it may be a realistic proxy for patch correctness for two reasons.
First, with the use of LLMs that are capable of generating hundreds of candidate patches, manually evaluating each generated patch for semantic equivalence can be prohibitively expensive (in the order of 28,000 per model configuration). Thus, evaluating candidate patches with the developer written tests serves as a scalable proxy for preserving functionality and defect-freedom. 
Secondly, semantic equivalence with user-provided fix may not be necessary, as there can be multiple, non-equivalent yet acceptable, fixes to developers in practice. Without knowledge of the detailed invariants of the project, it is difficult to determine if a particular ground truth fix is the only acceptable fix.

We augment the Defects4J dataset in three distinct ways:

\begin{itemize}[leftmargin=*]
\item We restrict the \dforjnltoedit{} dataset to consist of fixes that affect a {\it single method} body. Among 835 bugs in the Defects4J 2.0 dataset, 283 contain single-method bugs, i.e., bugs that can be fixed with single method changes. Fixes may include multiple lines, but are scoped to a singular function. Table~\ref{tab:dataset_statistics} contains a breakdown of the number of bugs per project.
 
We justify our decision to focus on single method bugs as methods generally define a unit of code that can be reviewed independently compared to isolated lines, or an entire file or repository.
Second, the input prompt for LLMs are restricted to only a few thousand tokens that may not suffice to capture the entire file or repository level information. APR approaches using the Defects4J dataset often restrict their dataset to contain only single hunk, or single line bugs\cite{alpharepair-22}. We do not make this restriction, and \Cref{tab:dataset_statistics} shows the average number of hunks for bugs in our dataset.

\item Second, to serve the \nlTofix{} problem, we augment the Defects4J dataset by pairing each bug with its corresponding issue metadata, including the issue title and description, that we scrape from GitHub, SVN and Jira.

\item Finally, upon close investigation of the buggy methods in the Defects4J dataset, we notice that as a side-effect to the bug patching process used by the benchmark creators, comments that appear in the \ic{POST\_FIX\_REVISION} also appear in the \ic{PRE\_FIX\_REVISION} of the code\footnote{https://github.com/rjust/defects4j/issues/477}. This means that comments related to the actual fix made by the developer, may appear in the \ic{PRE\_FIX\_REVISION} that we use as input to the LLMs. To avoid these comments providing hints about the solutions to the model, we remove all comments from the \ic{PRE\_FIX\_REVISION}. 

\end{itemize}

\subsection{Generative Pre-trained Transformers (GPT)}

Generative Pre-trained Transformers (GPT) are large-scale auto-regressive~\cite{johansen1995likelihood} generation models trained to predict the next token given a natural language prefix (prompt) context. The recent development of ultra-large-scale GPT models with billions of parameters has shown to exhibit emergent properties where they can perform tasks without finetuning~\cite{patel2022bidirectional, kojima2022large}. While asked to generate responses to a prompt, GPT models samples over tokens' probability distributions of one token at a time. To generate the most probable response (or multiple responses), these models perform different sampling, including temperature-based sampling, which manipulates the distribution of tokens,  controlling the diversity in the responses. A lower temperature typically results in less diversity, and a higher temperature otherwise. We use two temperate -- 0.2 and 0.8 throughout the experiments. 

To answer the defined research questions, we select three state-of-the-art GPT-based Large Language Models (LLM) that have shown strong capabilities on a variety of code generation tasks\footnote{At the time of submission, the authors did not have API access to the most recent state-of-the-art model, GPT-4~\cite{openai2023gpt4}}.

\textbf{Codex.} OpenAI's Codex, \completion is a language model specifically designed for code completion tasks. It is based on the GPT-3 architecture and has been fine-tuned on a large corpus of code from public repositories. Codex excels at generating syntactically correct code and has been shown to be highly effective for tasks involving code generation.

\textbf{Codex Edit Model.} The Codex edit model, \edit, is a version of Codex GPT-3 with editing capabilities. Given a code and instruction written in NL such as "Improve the runtime complexity of this function", the models edits the code to possibly satisfy the instruction.

\textbf{ChatGPT.} The recently released ChatGPT model (\ic{gpt-3.5-turbo}) is based on the pretrained GPT-3.5 model, which is further finetuned using Reinforcement Learning with Human Feedback (RLHF)~\cite{ouyang2022training}. While \ic{gpt-3.5-turbo} is not explicitly fine-tuned for code generation tasks, early evaluation  has demonstrated strong capabilities in several fields of science and engineering~\cite{qadir2022engineering, jiao2023chatgpt, qin2023chatgpt, kung2023performance} including understanding and generating code snippets~\cite{maddigan2023chat2vis, sobania2023analysis, surameery2023use}. ChatGPT's conversational nature allows it to excel in tasks that require both code generation and human-like interactions, allowing the use of advanced prompt structures that involve chain of thought\cite{wei2022chain} and reasoning extraction\cite{kojima2022large}.

\textbf{Embeddings.} OpenAI embedding models generate a high dimension vector representation of input strings. 

{Research shows that similarity in such high-dimension vector space translates to semantic similarity of strings.} 
Among many other applications of such representation, they can facilitate similarity analysis, searching, etc. In this work, we leverage the text-embedding-ada-002 model to generate the embedding of code and use embedding-based similarities to rank the patches.

\subsection{Prompting Framework}
\label{sec:prompting}
 Capabilities of LLMs are not fixed across all contexts, \ie if an LLM gets a question wrong, slightly changing the prompt by modifying the contents or format of the information given may yield different outcomes. There as several techniques to improve the accuracy and reliability of LLM output, these techniques are referred to as prompt engineering\cite{reynolds2021prompt}. In this paper we use standard prompting as well as two distinct strategies that have been shown to improve the performance of LLMs for complex tasks: 1) few-shot prompting and 2) reasoning extraction. These strategies are designed to help provide context and guidance to effectively solve a task while mitigating potential pitfalls associated with model-generated outputs.
 
 \subsubsection{Zero-Shot Prompting}
 Zero-shot prompting, also frequently referred to as \emph{standard prompting}, is the basic configuration of prompting a model with a task. The prompt does not include any examples of acceptable solutions and does not break down the problem into easier sub-problems. In this paper, our zero-shot, or standard, prompt is illustrated in \Cref{fig:overview}. It is composed of the issue title and description, along with the buggy code and an instruction to provide a fix to the code. 
 
 \subsubsection{Few-Shot Prompting}
 Few-shot prompting is a technique that involves presenting the model with a series of examples or demonstrations in order to guide its understanding of the task at hand. By providing the models with a few instances of similar tasks, along with their respective inputs and desired outputs, we guide the model output to the desired output, both in terms of functionality and format. This approach enables the models to adapt their responses based on the provided examples, leading to more accurate and coherent output. 
 For a given issue, we select another issue along with the buggy code and fixed code to represent as a shot. We select the example to be an issue where the example buggy code is closest to the target buggy code, using standard edit distance metric.

\subsubsection{Reasoning Extraction}
Reasoning extraction is a strategy that focuses on extracting the underlying rationale behind a specific task or problem~\cite{kojima2022large}. We apply this strategy to help the model comprehend the objective(s) and solution to the code fix task. In particular, we explicitly interact three times with the model with different queries. First, given the buggy code and issue report, we ask the model to localize the bug, then we ask to explain why the localized lines are buggy, finally we ask to fix the bug. ChatGPT’s conversational nature allows the natural the use of advanced prompt structures that maintain conversational context, like chain of thought reasoning and reasoning extraction. Therefore, we only use this prompt strategy with \gpt.

\subsection{Correctness of Generated Code}
Experiments are run in two phases: fix generation and fix validation. All validation experiments are run in a Docker container running Ubuntu 20.04.4 with Java version OpenJDK 1.8.0 for which we make the Docker image public. To generate candidate fixes we use the OpenAI API.  The rest of this section discusses details of patch validation and evaluation metrics:
\subsubsection{Patch Validation}
Each bug in the Defects4J dataset contains a \ic{PRE\_FIX\_REVISION} and \ic{POST\_FIX\_REVISION} version that represents the buggy/fixed versions of the code respectively. The two versions reflect the actual state of the project when the bug was discovered/fixed. To determine whether a generated fix is correct, we follow the following steps: 1) Check out the \ic{PRE\_FIX\_REVISION} version of the project 
2) Replace the original buggy function with the generated function and 3) Run the trigger and regression test(s) to determine if code containing the generated fix passes the tests or not. 

For each fix, the validation outcome can be either 1) Plausible: all bug reproducing tests and regression tests pass 2) Wrong: at least 1 of the trigger or regression tests fail or 3) Uncompilable.

\subsubsection{Evaluation metrics}
\label{sec:passk}

To measure the quality of a solution to the \nlTofix{}, we use the pass@k metric\footnote{we use pass@k and P@k interchangeably throughout the paper} introduced and widely used for evaluating LLMs for \nlTocode{} problems~\cite{codex_2021,google_llm_2021}.
Intuitively, given an unordered set of candidate fixes, the pass@k provides the likelihood of choosing a correct fix when given $k$ tries to sample from this set of candidate fixes.
In \nlTofix{} scenario, a fix is correct if it passes all the Trigger tests and Regression tests for the bug. Given $n$ as the number of samples generated, $k$ as the number of samples to estimate pass@k and $c$ is the number of correct samples in $n$, we use a the following formula for calculating pass@k defined by \cite{codex_2021}: 

\begin{equation}
    \text{pass@k} :=  \underset{\text{problems}}{\mathbb{E}}\left[1 - \frac{\binom{n - c}{k}}{\binom{n}{k}}\right]
\end{equation}

 For this paper, we generate $n \geq k$ samples, where $n=100$.
 Additionally, to answer RQ4, we also introduce a ranked pass@k, denoted r.pass@k (\Cref{sec:rq4}), to determine the number of cases where at least one code fix is correct in the top $k$ suggestions of a ranked list.

\section{Results}
\label{sec:results}

\begin{figure}[t]
  \centering
  \begin{subfigure}[t]{0.49\linewidth}
    \includegraphics[width=\textwidth]{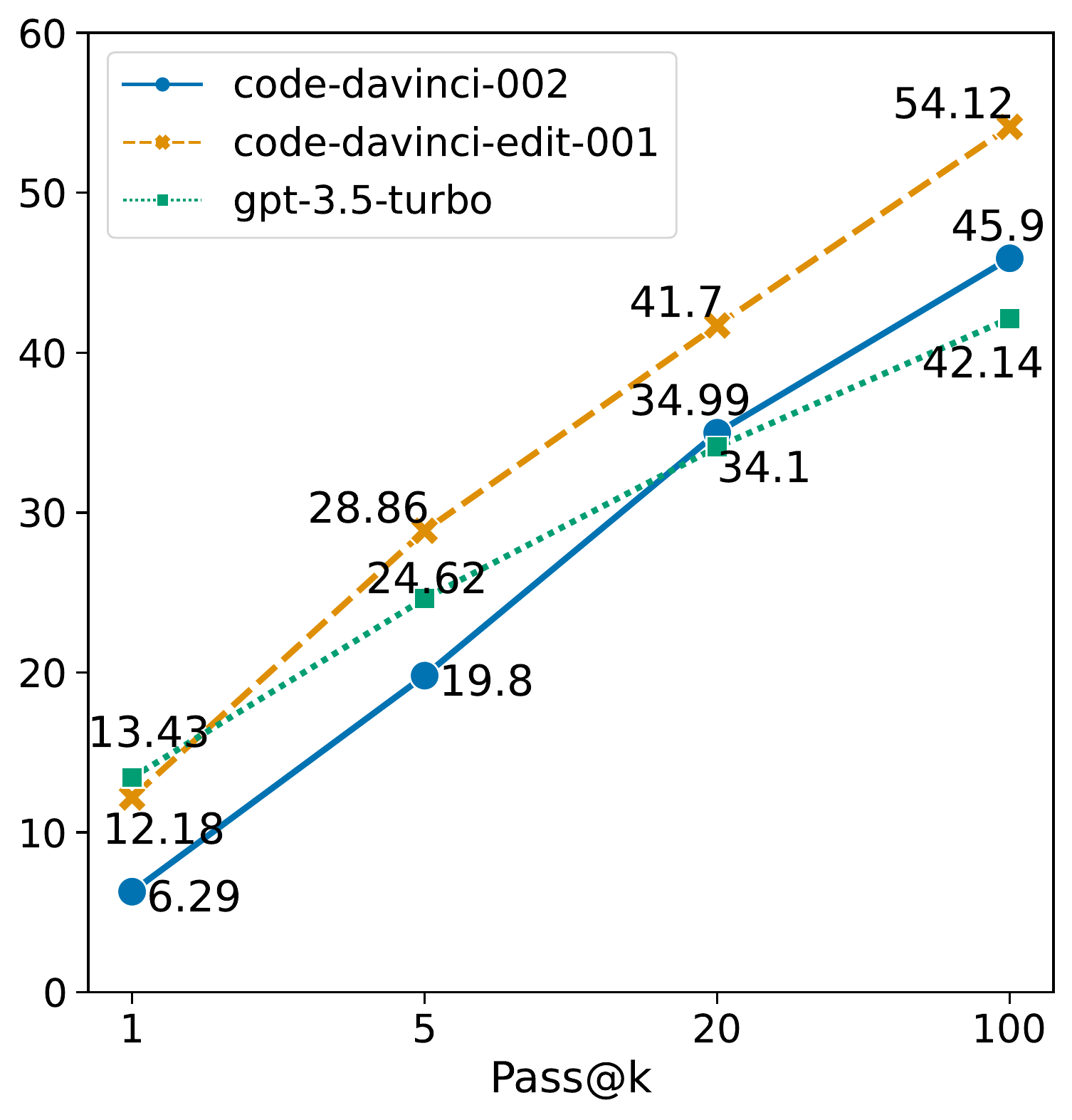}
    \caption{Temp=0.8}
      \label{fig:p-k-8}
  \end{subfigure}
  \begin{subfigure}[t]{0.49\linewidth}
    \includegraphics[width=\textwidth]{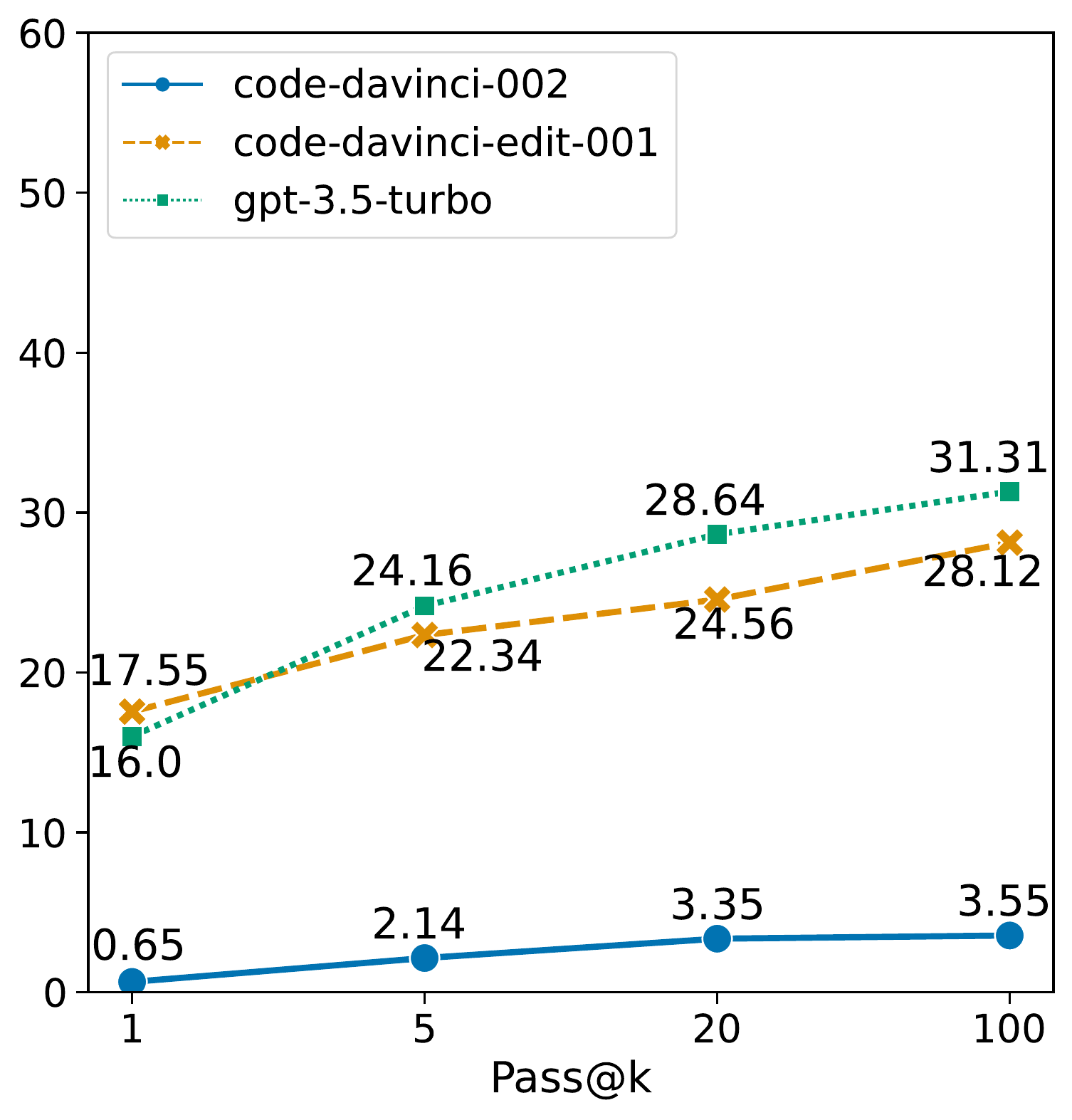}
    \caption{Temp=0.2}
    \label{fig:p-k-2}
  \end{subfigure}
\caption{Pass@k results for 0-shot setting.}
\end{figure}

\subsection{RQ1: \rqa}

For each model, we consider two settings with the zero-shot prompt: temp. at 0.2 and temp. at 0.8. For each setting we generate 100 candidate fixes for each of the 283 bugs and evaluate the correctness of each candidate against the trigger and relevant tests. Next, we calculate the pass@k using technique described in \Cref{sec:passk} and plot the results for temp. 0.8 in Figure~\ref{fig:p-k-8} and temp. 0.2 in Figure~\ref{fig:p-k-2}. 

At temp. 0.8, we see that the edit model,\edit, is the best performing model overall with a pass@100 at 54.12\%. \gpt achieves a slightly higher pass@1 (13.93\% with respect to 12.18 for the edit model), however at pass@20 and above, performance dips lower than of the completion model, \completion.  The completion model achieves the second highest pass@100 with 45.9\%, which is 8.22\% lower than the edit model. 

At temp. 0.2, we see precision improvements for pass@1 from both \edit and \gpt with 17.55\% and 16.0\% respectively. However, we see consistently lower precision of all models starting at pass@5 through pass@100. Most notably, the precision of \completion is much lower at temp. 0.2 with just 0.65\% pass@1 and 3.55\% pass@100. We tested the precision of the model in two separate runs, and noticed consistently poor performance at this setting.

To better understand the pass@k accuracy per model, we extracted high level statistics about the code generated by each model for both temp. configurations. Table~\ref{tab:patch-stats} contains the average percentage of duplicate code candidates generated per bug as well as the average percentage of candidates that compile, pass on the regression tests, and pass on both the regression and trigger tests (plausible) across bugs. 

From Table~\ref{tab:patch-stats} we can see that the number of duplicates generated increases drastically when the temp. is decreased, which is expected as model behavior is more deterministic at lower temperatures. For example, \edit, which is optimized for code edits, generates more than 90\% duplicates at 0.2 but only 26\% at 0.8. 

Overall, the percentage of generated code that compiles varies significantly across models. We observe that only 4.66\% of code generated by \completion at temp. 0.2 compiles, which explains the extremely low accuracy seen in Figure~\ref{fig:p-k-2}. However, at temp. 0.8 the compilation rate increases significantly, between 35.9\% and 74.3\%. Looking to related work, on a different subset of the Defects4J dataset, SOTA neural APR techniques generate patches with 15\% to 28\% compilation rates in top 100 \cite{alpharepair-22}\cite{lutellier2020coconut}\cite{jiang2021cure}. Although the APR setting is different from that of nl2fix, the trigger and relevant tests \emph{are not} hidden from the APR setting, we observe that using the entire method as input to the LLMs have an advantage over generating a higher proportion of compilable patches.

% \usepackage{booktabs}
\begin{table}[t!]
\centering
\caption{Average candiate patch statistics across bugs. }
\label{tab:patch-stats}
\resizebox{\linewidth}{!}%
{
\begin{tabular}{@{}lcc|cc|cc@{}}
\toprule
 & \multicolumn{2}{c}{\textbf{\edit}} & \multicolumn{2}{c}{\textbf{\completion}} & \multicolumn{2}{c}{\textbf{\gpt}} \\ 
Temp. & 0.2 & 0.8 & 0.2 & 0.8 & 0.2 & 0.8 \\ \midrule
\textbf{Duplicate} & \textbf{90.7}\% & 26\% & 87.9\% & \textbf{35.2}\% & 51.8\% & 8.41\% \\ \midrule
\textbf{Compile} & \textbf{74.3}\% & 54.8\% & 4.66\% & 35.9\% & 57.2\% & \textbf{60.4}\% \\
\textbf{Regression} & \textbf{51.2}\% & \textbf{41.7}\% & 3.23\% & 20.19\% & 25.1\% & 33.0\% \\
\textbf{Plausible} & \textbf{20.9}\% & 12.0\% & 0.65\% & 6.0\% & 16.0\% & \textbf{13.4}\% \\ \bottomrule \\
\end{tabular}
}
\end{table}

Both \gpt and \edit achieve a higher precision for pass@100 in the temp. 0.8 setting, compared to 0.2 (Figure~\ref{fig:p-k-2}). However, in table~\ref{tab:patch-stats} we see that the average percentage of plausible patches is higher in the 0.2 setting. While this appears counter intuitive, the presence of high number of duplicates per patch modulates the calculated pass@k across bugs. In other words, a model may have high confidence for a small number of bugs and generate a high percentage of plausible patches for that bug.

\RS{1.1}{Given only an NL decription of a bug, all three LLMs are able to generate plausible fixes for a modest number of bugs in the dataset, with pass@1 of at most 6.29\% -- 17.55\% and pass@100 of at most 42.19\% -- 54.12\%. In the 0-shot setting, \edit achieves the overall highest accuracy compared to both \completion and \gpt. }

We report the number of bugs with plausible fixes for each project in \Cref{tab:patch-per-project}. At a high level, we observe that the distribution of plausible fixes generated from each model is distributed across every project. In general for every project, the three models generate plausible fixes for a similar percentage of bugs. 
There are a few notable exceptions, for example, the Collections project, which only contains 1 bug, was only correctly fixed by \edit. 
\gpt also has lower accuracy on the Mockito and JacksonDatabind projects only generating plausible fixes for 1/21 and 7/67 bugs respectively, compared to 6/21 and at least 22/67 for the other two models. However, on the Codec project, \gpt generates plausible fixes for two more bugs than the other two models. Overall, the total bugs patched by each model aligns with the pass@100 metrics seen in \Cref{fig:p-k-8}.

\begin{table}[h]
\centering
\caption{Bugs with plausible patches, by project. Bugs with plausible fix / total number of bugs for that project.  }
\resizebox{0.81\linewidth}{!}%
{
\begin{tabular}{lccc}
\toprule
\multicolumn{1}{l}{\textbf{Approach}} & \ic{davinci-002} & \ic{edit-001} & \ic{gpt-3.5} \\ \midrule
\rowcolor[HTML]{RGBA(130, 130, 130, 0.08)} 
Chart & 6/6 & 6/6 & 5/6 \\
\rowcolor[HTML]{RGBA(130, 130, 130, 0.08)} 
Cli & 13/28 & 16/28 & 13/28 \\
\rowcolor[HTML]{RGBA(130, 130, 130, 0.08)} 
Codec & 7/11 & 7/11 & 9/11 \\
\rowcolor[HTML]{RGBA(130, 130, 130, 0.08)} 
Collections & 0/1 & {\color[HTML]{3166FF} 1/1} & 0/1 \\
\rowcolor[HTML]{RGBA(130, 130, 130, 0.08)} 
Compress & 15/36 & 19/36 & 18/38 \\
\rowcolor[HTML]{RGBA(130, 130, 130, 0.08)} 
Csv & 10/12 & 8/12 & 9/12 \\
\rowcolor[HTML]{RGBA(130, 130, 130, 0.08)} 
JacksonCore & 8/13 & 9/13 & 5/13 \\
\rowcolor[HTML]{RGBA(130, 130, 130, 0.08)} 
JacksonDatabind & 22/67 & 27/67 & {\color[HTML]{CB0000} 7/67} \\
\rowcolor[HTML]{RGBA(130, 130, 130, 0.08)} 
JacksonXml & 1/5 & 2/5 & 2/5 \\
\rowcolor[HTML]{RGBA(130, 130, 130, 0.08)} 
JxPath & 3/10 & 5/10 & 4/10 \\
\rowcolor[HTML]{RGBA(130, 130, 130, 0.08)} 
Math & 38/73 & 45/73 & 45/73 \\
\rowcolor[HTML]{RGBA(130, 130, 130, 0.08)} 
Mockito & 6/21 & 6/21 & {\color[HTML]{CB0000} 1/21} \\ \midrule
\multicolumn{1}{l}{\textbf{Total}} & \textbf{129/283} & \textbf{151/283} & \textbf{118/283} \\ \bottomrule \\
\end{tabular}}

\label{tab:patch-per-project}
\end{table}

 \begin{figure}[h]
\includegraphics[width=0.8\linewidth]{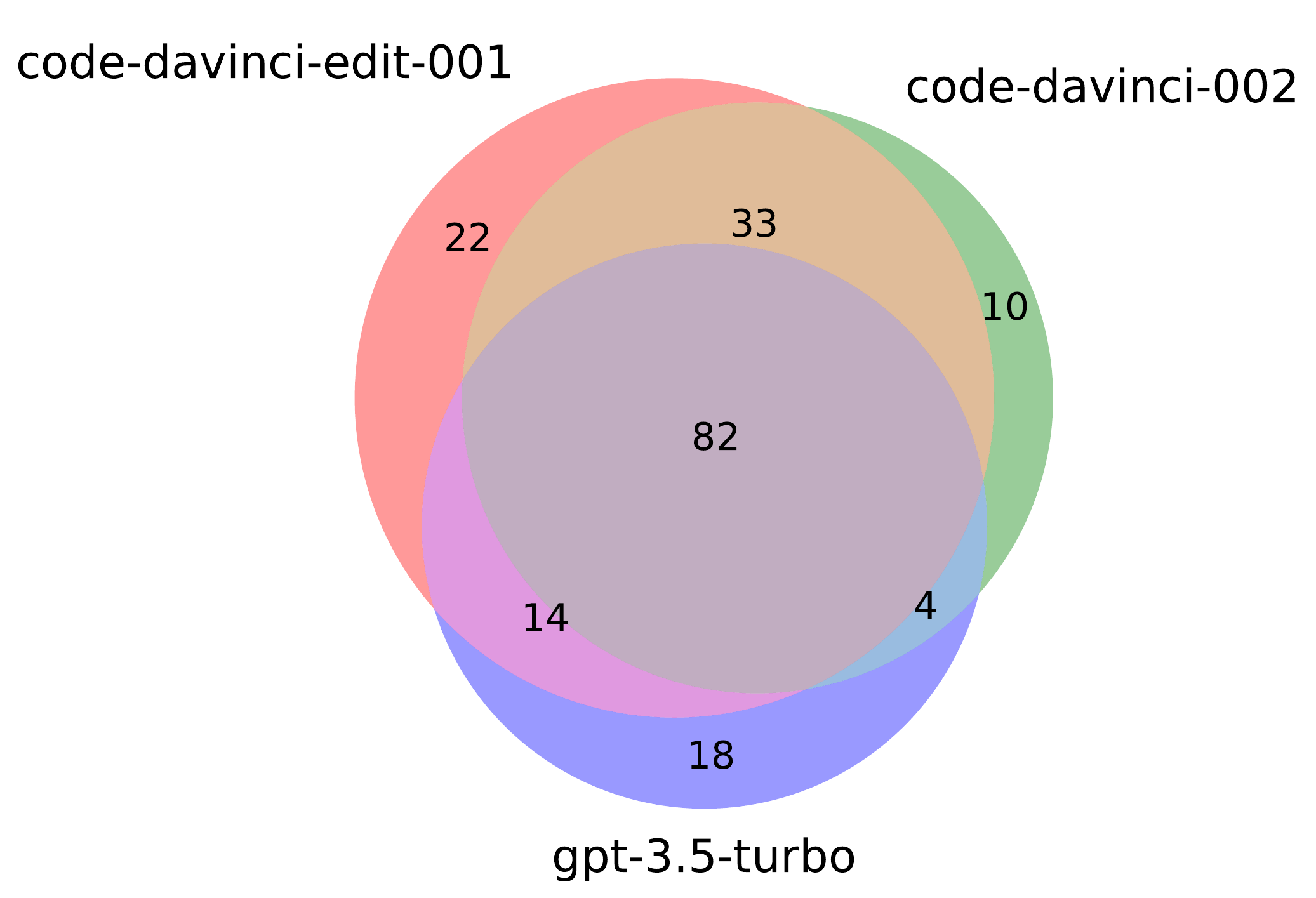}
\caption{Overlap between bugs with plausible patches for each LLM in the 0-shot temp 0.8 setting.}
\label{fig:bug-overlap}
\end{figure}

While two models may patch the same number of bugs for a project, the exact bugs that are patched may be different. Figure~\ref{fig:bug-overlap} shows the overlap of the number of bugs each model is able to generate plausible patches for. All three models are able to generate plausible patches for the same 28\% (82 of 283) of the bugs. However, we observe that when combined together, the three models can generate plausible patches for 64\% (183 of 283) of the dataset. Each model has a unique subset of bugs that the other two models are not able to generate plausible patches for: 22  unique bugs by \edit, 18 by \gpt, and 10 by \completion. While some bugs may be easier to fix for certain models, this does not appear to be a consistent artifact of the project that the bug originates from, as observed from \Cref{tab:patch-per-project}.

\RS{1.2}{28\% (82/283) of the bugs in the dataset can be fixed by all three LLMs. When combined together, the three LLMs can generate plausible patches for 64\% (183/283) of the bugs in the dataset.}

\subsection{RQ2: \rqb}
\label{sec:rq2}

\begin{figure*}[t!]
    \centering
\begin{subfigure}[t]{.30\textwidth}
    \centering
    \includegraphics[width=\linewidth]{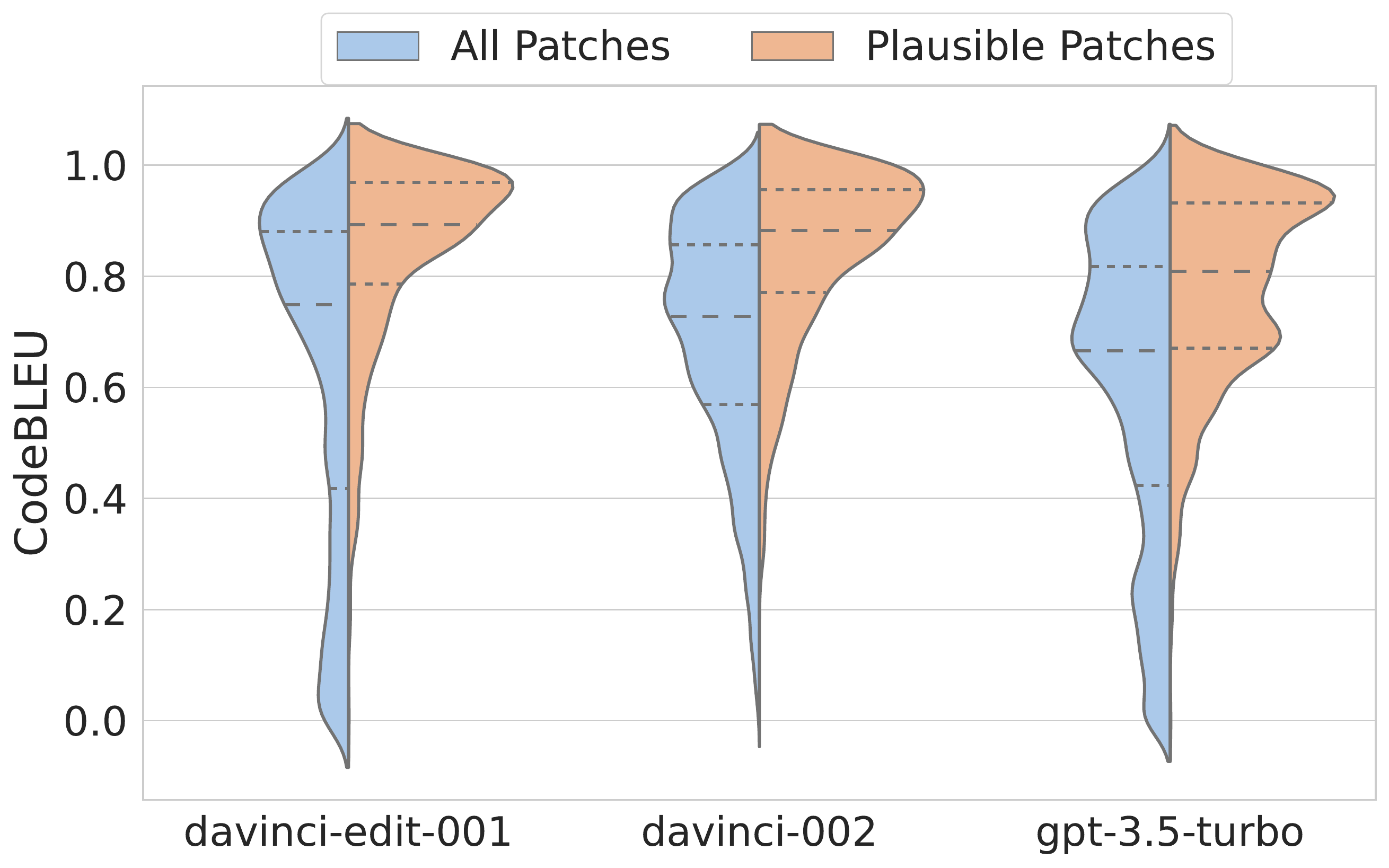}
    \caption{With the actual fixed code.}
    \label{fig:patch_similarity_from_fixed}
\end{subfigure}%
\hfill
\begin{subfigure}[t]{.30\textwidth}
    \centering
    \includegraphics[width=\linewidth]{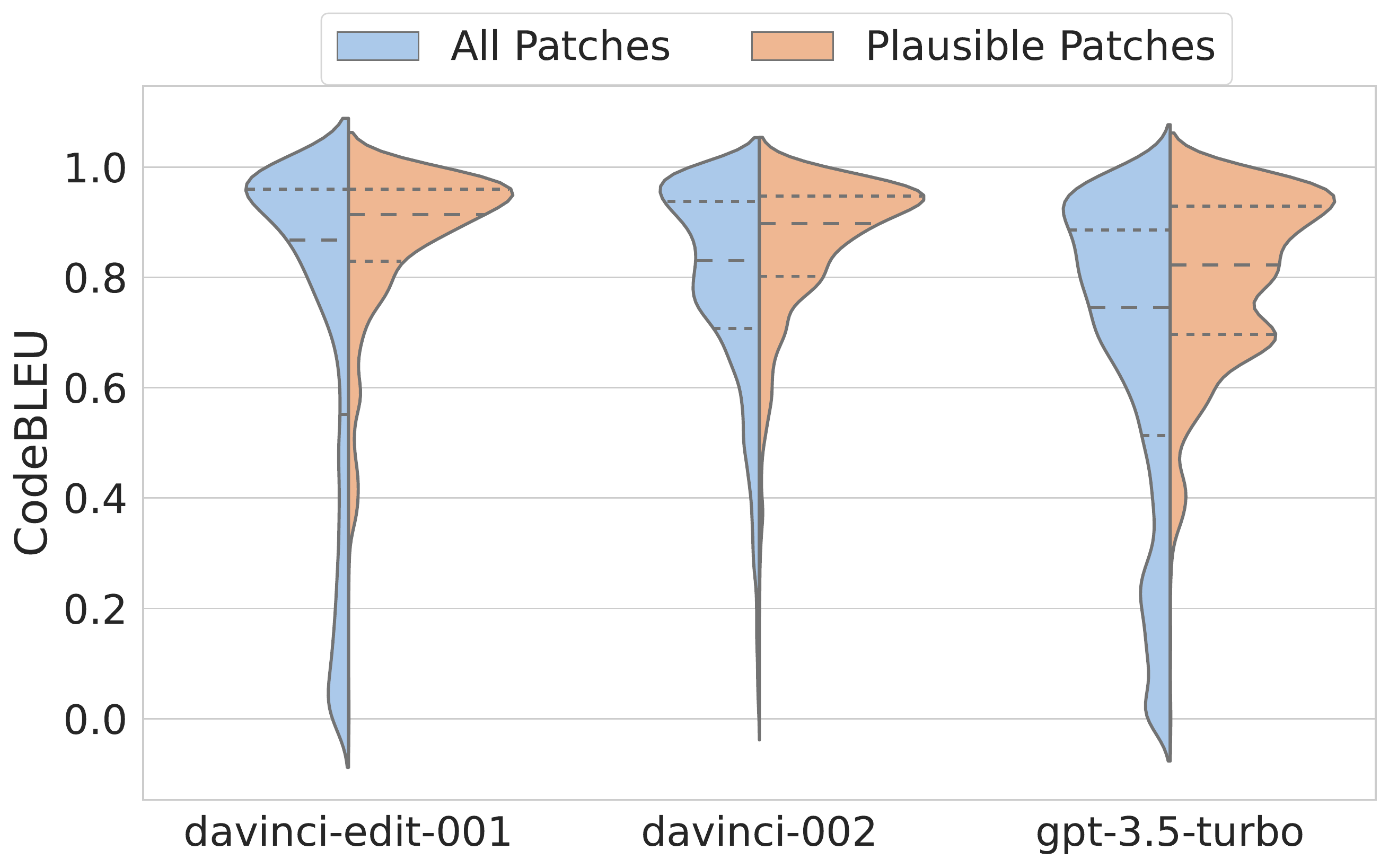}
    \caption{With the Buggy code.}
    \label{fig:patch_similarity_from_buggy}
\end{subfigure}%
\hfill
\begin{subfigure}[t]{.30\textwidth}
    \centering
    \includegraphics[width=\linewidth]{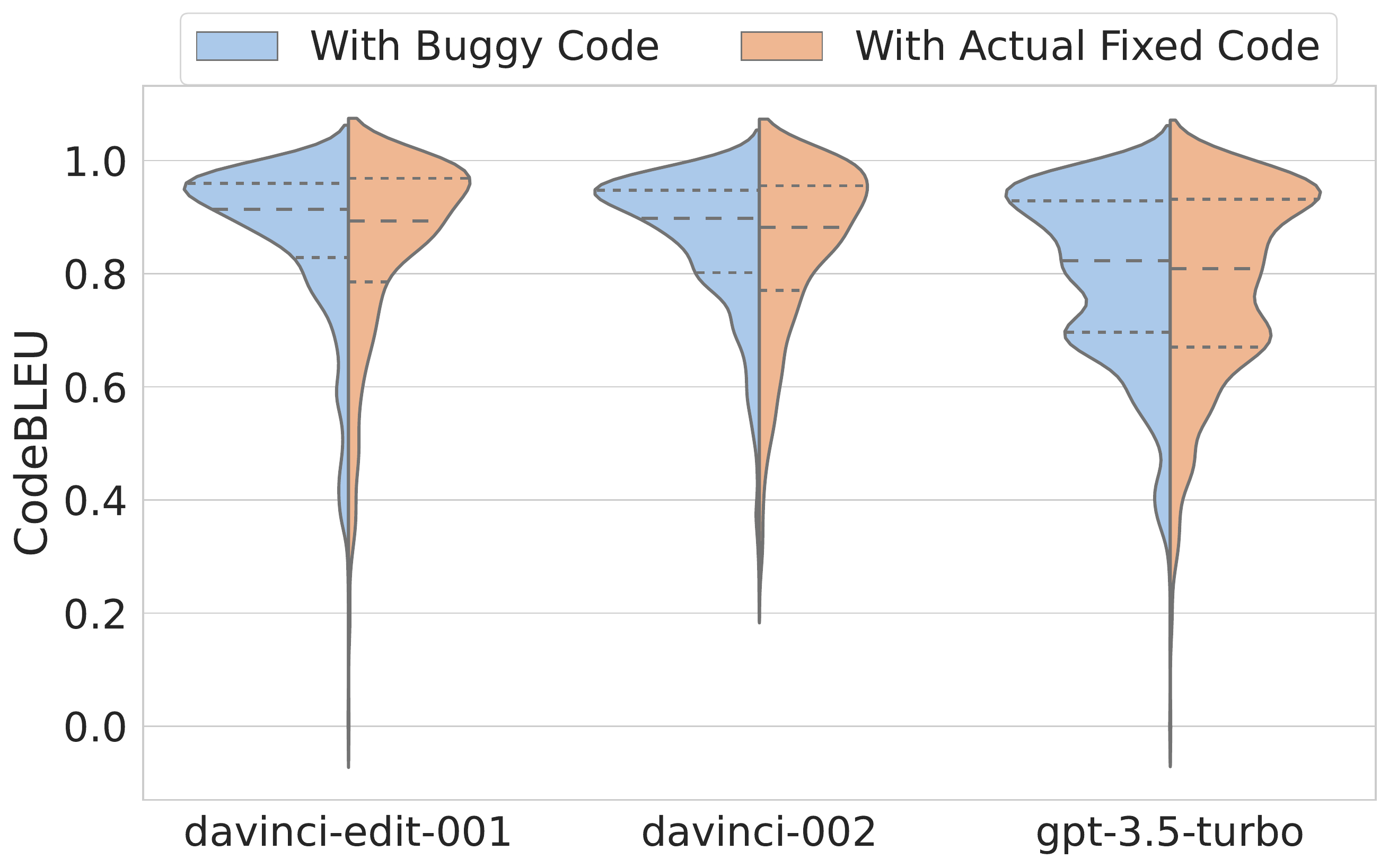}
    \caption{Plausible patches with input buggy code and actual fixed code.}
    \label{fig:plaus_patch_similarity_from_buggy_and_fixed}
\end{subfigure}
\caption{Code similarity analysis of generated patches by different models. We analyze the CodeBLEU similarity here.}
\label{fig:characteristic_violin}
\end{figure*}

To answer this RQ, we select the best-performing configuration from RQ1 (temp. 0.8) to better understand the nature of the code generated by the LLMs. We study the characteristics of the patches generated by different models. 
To understand the characteristics of different patches, we study the similarity of those patches \wrt, the buggy code (present as part of the input) and the actual fixed code.
We use CodeBLEU~\cite{ren2020codebleu} as the representative similarity measurement. Given two code $c_1$ and $c_2$, CodeBLEU is defined as 
$\alpha*B + \beta*W + \gamma*S + \delta*D$, where $B$, $W$, $S$, $D$ are BLEU score, Keywords BLEU score, Syntax match score, and Dataflow match score, respectively between $c_1$, and $c_2$, and $\alpha$, $\beta$, $\gamma$, $\delta$ are weighting constants typically all set to 0.25~\footnote{\href{https://github.com/microsoft/CodeXGLUE/tree/main/Code-Code/code-to-code-trans/evaluator/CodeBLEU}{Microsoft's CodeBLEU implementation}}. We choose CodeBLEU for this research question since it considers the syntax and semantic match between code, in addition to the lexical match.

\begin{figure*}[t!]
    \centering
    \includegraphics[width=.80\textwidth]{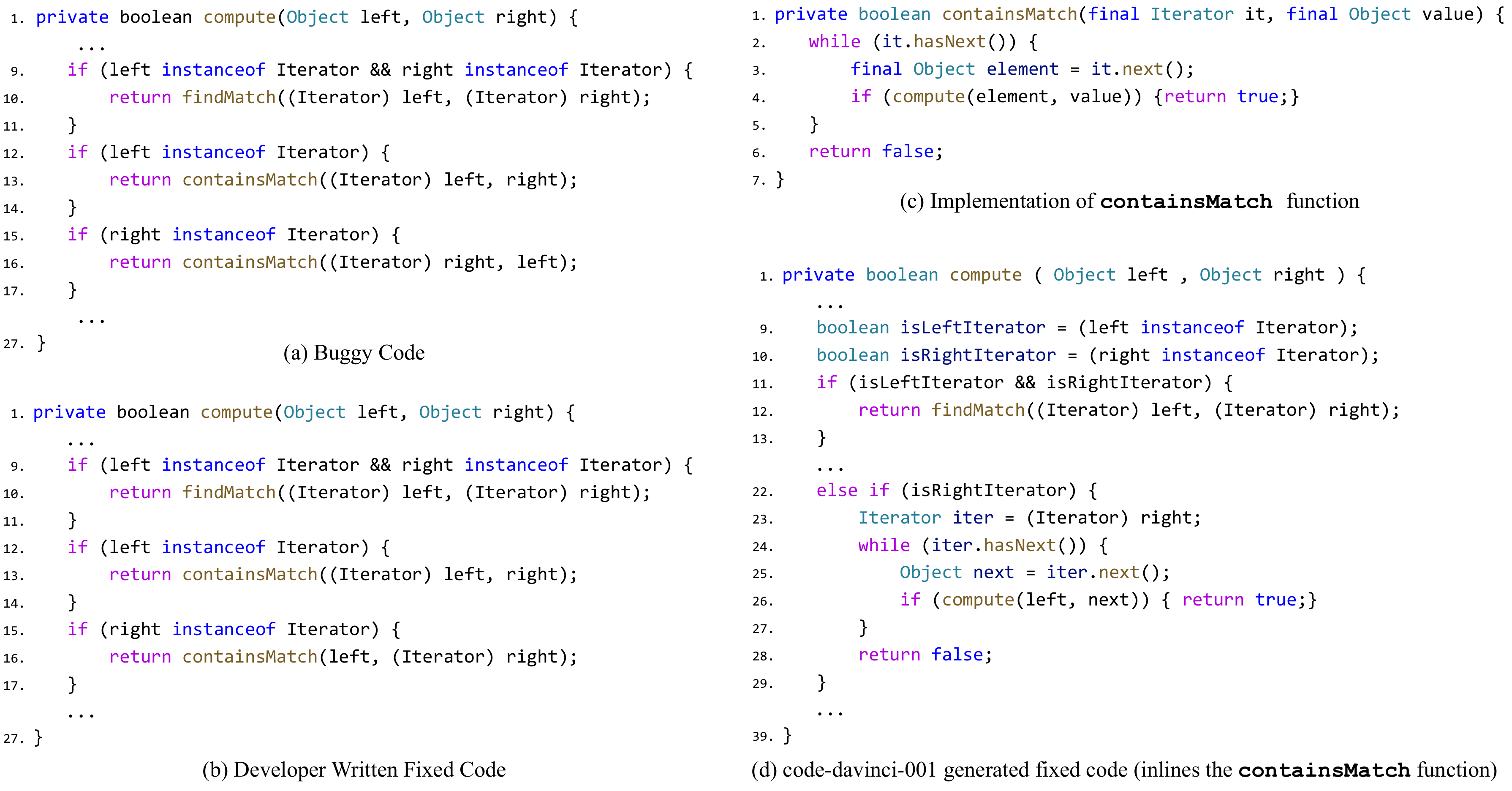}
    \caption{An example showing the contrast between actual fixed code and model-generated plausible patch for Bug id 20 of JxPath project. Even though the generated patch passed does not exactly match the ground truth fix, it passed all the regression tests and the trigger test, making it a semantic equivalent of the actual fix. 
    }
    \label{fig:semantic-equivalence}
\end{figure*}

\Cref{fig:patch_similarity_from_fixed} shows the similarity of patches with the actual fixed code. 
Across all three models we have studied in this paper, the patches that passed both the regression test and the bug revealing trigger test (\ie plausible patch) exhibit higher CodeBLEU with the actual fixed code compared to the patches that are not plausible. Such a result is expected since the plausible patches pass the whole test suite; the plausible patch should be, in theory, a semantic equivalent of the actual fix exhibiting higher CodeBLEU. Interestingly, the plausible patches generated by the \gpt model exhibit higher variability regarding CodeBLEU similarity with the actual patch. The Inter-Quartile Range (IQR) of CodeBLEU between plausible patches and actual fixed code is 0.18 and 0.19, respectively, for \edit and \completion. In contrast, for \gpt, the IQR is 0.26. In addition, for the \edit and \completion models, the kurtosis are 2.62 and 1.12, respectively, signifying the distributions are more centered, whereas \gpt models kurtosis is 0.41, signifying diverse generation capability, as evident from \cref{fig:patch_similarity_from_fixed}.

\lessons{\gpt generates more diverse patches as compared to other models.}

Further, we analyze how similar the generated patches are \wrt, the buggy code. \Cref{fig:patch_similarity_from_buggy} shows the distribution of CodeBLEU of different types of patches across different models. Across all three models, interestingly, we observe that plausible patches exhibit higher CodeBLEU with the buggy code than their non-plausible counterparts. We conjecture that when models make extensive modifications of the input buggy code, the resultant code are infested with different problems causing them to fail compilation, regression tests, and the trigger test (see \Cref{tab:patch-stats}). We conjecture that LLMs would make a more significant impact on the \nlTofix{} problem if we had the option to control the deviation from the input buggy code. Nevertheless, the observation that plausible patches exhibit higher similarity with buggy code opens up a new possibility of ranking the LLM-generated code based on their similarity with buggy code, which we will investigate in detail in the following research question. 

\lessons{Plausible patches exhibit higher similarity with the buggy code than non-plausible patches.}

\begin{table}[t]
    \centering
    \caption{Statistics of bugs and patches where models generate patches which exactly matches the ground truth fix}
    \label{tab:bugs_copy_stat}
    \resizebox{\linewidth}{!}%
    {
    \begin{tabular}{lcccc}
        \toprule
          \multirow{2}{*}{\textbf{Model}} &  \textbf{\# Fixed} &  \textbf{\# EM}$^*$ &  \textbf{\# Plausible} &  \textbf{\# EM}\\
          & \textbf{Bugs} & \textbf{Bugs} & \textbf{Patches} & \textbf{Patches} \\
        \midrule
          \edit &  151 &  32 &  1724 &  146 \\
          \completion &       129 &  24 &  1182 &  120 \\
          \gpt &     118 &  11 &  2356 &  73 \\
        \bottomrule
    \end{tabular}
    }

    {\footnotesize $^*$ EM : Exact Match ignoring whitespaces}
\end{table}

There is a consensus among academics and practitioners that LLMs such as the davinci or gpt-3.5 models were pretrained on virtually all open source. As such, it is not unexpected that these LLMs have already observed the methods in our dataset during their respective pretraining. This raises a question, how much do the LLMs memorize from their pretraining~\cite{mccoy2021much, carlini2022quantifying, tirumala2022memorization}? 
Unfortunately, there is no good way to measure such unless we know what LLMs' pretraining data is. 
Regardless, to qualitatively understand the generated patches, we investigate the CodeBLEU similarity of the patches with the buggy code (which was available in the input to LLM) and the actual fixed code. Across all three models, generated patches exhibit slightly higher similarity with the buggy code than the actual fixed code (see \Cref{fig:plaus_patch_similarity_from_buggy_and_fixed}). Such difference is statistically significant by one-sided Wilcoxon sign rank test with p-values of $1.6*10^{-9}$, $9.9*10^{-5}$, $2.7*10^{-14}$ for \edit, \completion, and \gpt, respectively.

\Cref{tab:bugs_copy_stat} shows summary statistics of patches that match the dataset's ground truth fix exactly. The \edit model correctly generated at least one patch for 151 bugs, among which the patch for 32 bugs exactly matches the ground truth. For the \completion model, such number is 24 out of 129 and 11 out of 118 for the \gpt. Only 8.5\% (146 out of 1724), 10.1\% (120 out of 1182), and 3\% (73 out of 2356) patches of the plausible patches are an exact match with the ground truth for \edit, \completion, and \gpt, respectively. These results show that most of the plausible patches are syntactically distinct from the ground truth fix.

In \Cref{fig:semantic-equivalence}, we show one of the plausible patches generated by \gpt for the example in \Cref{fig:overview}. The buggy code misplaced the arguments of {\tt containsMatch} function in line 16(\cref{fig:semantic-equivalence}(a)), while the developer-written patch fixed the error by putting the arguments in correct positions (line 16 in \cref{fig:semantic-equivalence}(b)). \Cref{fig:semantic-equivalence}(d) shows a fixed code generated by the \gpt model, which is semantic equivalent of the developer-written code. In fact, LLM generated fix actually inlines the implementation of {\tt containsMatch} (shown in \cref{fig:semantic-equivalence}(c)) function into the context (line 23-28 in \cref{fig:semantic-equivalence}(d)). In addition, the LLM-generated patch refactors the code by extracting two variables corresponding to two boolean expressions used in the original code, making the resultant code more readable.  
We observe that even though arguably LLMs already had seen everything open source, they explore new variations of code when applied to the NL2Fix problem. 

\lessons{LLMs may not always memorize code from their pretraining. They explore new code while trying to generate patches.} 

\RS{2}{Across all three models we studied, the plausible patches exhibit higher similarity with input buggy code than non-plausible patches with differences in the median similarity of $\sim$0.15. The plausible patches also exhibit higher similarity with the buggy code than the actual fixed code with statistical significance (p-value $<<$ 0.05). The plausible patches exactly match the developer-written patch for up to 10\% of the  cases.}

\begin{table}[t]
\centering
\caption{Accuracy of different prompt configurations. }
\resizebox{0.95\linewidth}{!}%
{
\begin{tabular}{lllll} \toprule
Model & Prompt & P@1 & P@5 & P@100 \\ \midrule
\multirow{3}{*}{\edit}& Issue Title & 4.73 & 15.39 & 44.12 \\
& 0-shot & 12.18 & 28.86 & 54.12 \\
& 1-shot  & 4.73 & 12.75 & 30.24 \\
 \midrule
\multirow{4}{*}{\gpt} & Issue Title & 8.06 & 17.69  & 37.72 \\
& 0-shot & 13.43 & 24.62 & 42.14\\
& 1-shot  & \textbf{16.22} & \textbf{31.93} & \textbf{56.93} \\

& $^*$R.E. & 15.44 & 25.72 & 47.68\\ \midrule
\multirow{3}{*}{\completion} & Issue Title & 0.25 & 1.22 & 14.94 \\ 
& 0-shot & 6.29 & 19.8 & 45.91 \\
& 1-shot  & 3.92 & 14.34 & 40.09 \\
\bottomrule 
\end{tabular}}

{\footnotesize $^*$RE: Reasoning Extraction}
\label{tab:prompts}
\end{table}

\subsection{RQ3: \rqc}

To determine what information is needed to generate plausible fixes, we use different prompting techniques that provide different levels of information to each LLM and evaluate the pass@k metrics for each approach. In RQ1 and RQ2, we used a basic 0-shot prompt, containing the issue title and description, as described in \Cref{sec:prompting}. 

\begin{figure}[h]
\includegraphics[width=0.97\linewidth]{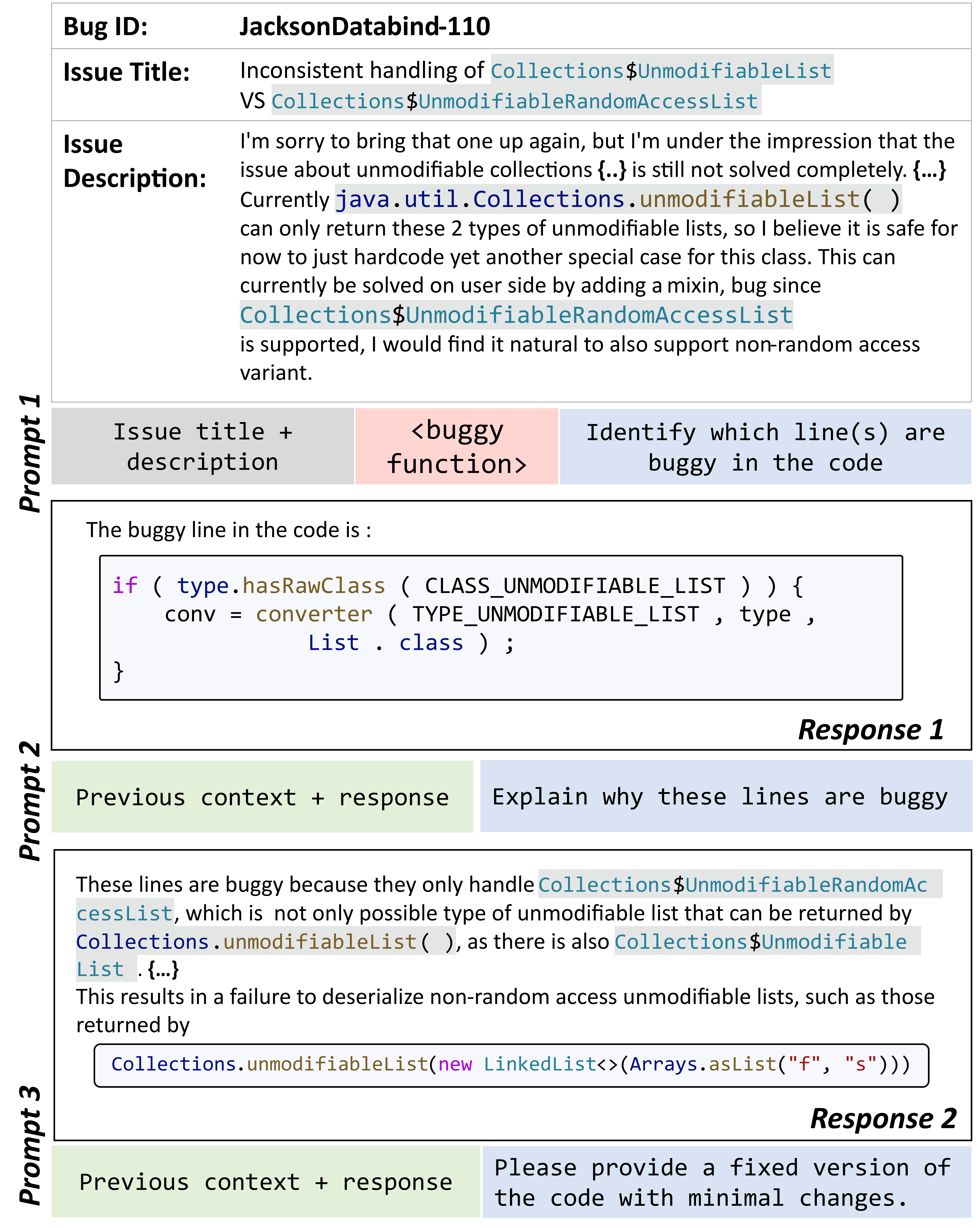}
\caption{A bug correctly patched using Reasoning Extraction.}
\label{fig:re}
\end{figure}

Table~\ref{tab:prompts} shows how changes to this basic prompt impacts pass@k for the following prompts: 
\begin{enumerate}[leftmargin=*]
\item 0-shot is the standard prompt structure used in RQ1-2, as represented in \Cref{fig:overview}.
\item 'Issue Title' indicates removing the issue description and preserving the rest of the prompt structure.
\item 1-shot adds an example of an issue, the corresponding buggy method, and the correct fix for the model to learn from. Then the rest of the prompt preserves the original format. The example chosen is based on its similarity to the code to be fixed. Details on this selection can be found in \Cref{sec:prompting}
\item For Reasoning Extraction, we modify the prompt that asks \gpt to break down the problem into two steps: 1) localizing the buggy lines in the original method and 2) explaining why these lines contain a bug before asking for the fix. An example is shown in \Cref{fig:re} 
\end{enumerate}

Across all three models, we observe that compared to using the issue title \emph{and} the description (0-shot) the issue title alone is not sufficient to achieve comparable P@k, with a decrease of close to 5\% -- 8\%  pass@1 and 5\% -- 30\% Pass@100. 
This indicates that the issue description contains valuable information for the model to generate plausible fixes. 

\lessons{Issue descriptions provide LLMs with helpful context to solve the NL2Fix problem.}

For \edit and \completion, we observe that adding examples in the prompt for 1-shot approach  does not improve performance, and in  fact decreases all pass@k compared to 0-shot attempts.  With the 1-shot setting, the average percentage of patches that compile for \edit falls from 54.8\% (\Cref{tab:patch-stats}) in the 0-shot setting to 24.7\%. Taking a closer look at the patches generated in the 1-shot attempt by \edit, we see several patches contain code from the 1-shot example in the prompt. For example, the original buggy method for Issue ID \ic{MATH-58} has the following method signature: \linecode{public double[] fit() \{...\}}. The examples used in the 1-shot setting has the function signature \linecode{private boolean isShortOption (String token) \{..\}}, and six of the generated patches edit the \\\linecode{isShortOption()} 
 function, instead of the buggy \linecode{fit()} function. 

This indicates that the model attempted to edit the target function using code from the example.  The \edit model expects an input code and a set of instructions for the edit. While OpenAI does not make the technical details of how \ic{code-davinci-} \ic{edit-001} was created, the observed behavior may indicate that this model was not fine-tuned with k-shot instructions, which may explain the degradation in model performance. Further, the context window, \ie the max number of tokens that can be used in the prompt, is much smaller for \edit, around 3000 tokens, compared to around 8000 for \completion and around 4000 for \gpt. Therefore, several examples are truncated to fit the context window to ensure that the target issue is still present in the prompt, leading to malformed method bodies. 

With the 1-shot setting, the average percentage of patches that compile for \completion falls from 35.9\% (\Cref{tab:patch-stats}) in the 0-shot setting to 20.5\%. When looking at the generated patches, we notice a large number are incomplete generations, \ie the code does not compile because the generated function is not syntactically correct. Examples of this include missing closing curly braces (\}), missing return statements, and completions that stop halfway through the target function. Note that there are two known ways of stopping the completion task \textemdash (a) by setting maximum length, (b) by setting specific STOP words. We set the maximum length to 750 tokens. In addition, we appended the method signature of the fixed method at the end of the prompt so that the completion model only needs to complete the body of the fixed method. 

\lessons{Information from examples used in 1-shot prompting do not help pass@k accuracy for \edit and \\\completion for the NL2Fix problem.}

On the other hand, compared to 0-shot attempts, pass@k for \gpt  improves for both 1-shot and reasoning extraction prompts. For example, pass@1 improves from 13.43\% to 16.22\% in the 1-shot setting and 15.44\% using reasoning extraction. \Cref{fig:re} shows an example of the interaction in the reasoning extraction prompt configuration. The example is from the JacksonDatabind project\footnote{https://github.com/fasterxml/jackson-databind/issues/2265}. When asked to identify the lines of code where the bug exists, \gpt returns the correct defect region in response 1. Then, prompt 2 appends the original prompt1 and response 1 as part of the context for prompt 2, with additional instructions to explain why the identified lines of code contain a bug. In response 3 \gpt explains the issue and extracts a sample input that the code would fail on and the corresponding error, from the issue description. In the final prompt, we append all inputs and responses to request a final fixed version of the buggy function. This is an example that \gpt is able to generate a plausible patch for in the reasoning extraction setting, but not the 0-shot setting. 

\lessons{Both Reasoning Extraction and 1-shot prompting approaches help \gpt to produce a higher number of plausible patches and increase pass@k.}

However, in the 1-shot setting \gpt is able to generate correct patches for 56 new bugs, and loses the ability to patch 14 bugs from the 0-shot setting. Using Reasoning Extraction, \gpt can generate correct patches for 34 new bugs, but loses the ability to patch 18 bugs  from the 0-shot setting. Compared to all approaches pooled together, \gpt in the reasoning extraction setting can only uniquely patch 4 bugs. Looking at the information contained in the issue descriptions for each of these examples, we observe \gpt is able to correctly localize buggy lines and reason about why they are buggy, but only with help from the context in the issue description. See \Cref{fig:re} for an example. While these prompting techniques do boost aggregate performance metrics, they also may degrade on a subset of the bugs in the dataset.

\RS{3}{Issue descriptions provide helpful context for solving the NL2fix problem. Prompting techniques that provide examples, \ie few-shot prompting, and break down the task, \ie reasoning extraction, significantly improve accuracy of \gpt on aggregate metrics like pass@k, however performance may degrade on certain subsets of the dataset and does not guarantee solutions over standard prompting. }

\subsection{RQ4: \rqd}
\label{sec:rq4}

Recall, given an unordered set of $n$ candidate solutions with $c$ correct solutions for a given problem (a bug in our case), the pass@k metric refers to the likelihood of picking at least one correct solution within $k$ tries. 
Such a statistics is useful for evaluating language models, but does not readily  provide an useful real-world recommender system that proposes a small number candidate fixes (upto say $k = 5$)  deterministically to a user.
For instance, for $n = 100$ and $k = 5$, there are $n \choose k$ > $75$ million ways to choose $5$ solutions from $100$ samples.
For a practical tool for \nlTofix{}, we would like to develop a ($i$) a deterministic way to {\it rank} the suggestions and present the top $k$ ranked suggestions to a user, and ($ii$) retain a high accuracy closer to the average pass@k metric.

In this section, we leverage LLMs to propose a simple and generic {\it ranking} strategy that helps realize the two objectives ($i$) and ($ii$) above. 
In particular, inspired by  our findings from RQ2, we explore if using a similarity between the embeddings of the input buggy function and the generated patches can identify plausible patches.
We generate embeddings for all buggy functions and corresponding patches using the \ic{text-embedding-ada-002} embedding model from OpenAI, see \Cref{sec:method} for details. We compute a cosine similarity between the embeddings for every pair of buggy code, and corresponding candidate patch.  We use this score to prune away patches with similarity scores lower than the median (0.95).  We fix this number across models for consistency. Then, to avoid ranking patches with extremely high similarity scores, \eg 1.0 , we rank the patches in order of lowest cosine similarity (starting at 0.95) to highest. Based on observations from RQ2, patches with lower scores are more likely to belong to the distribution of wrong or uncompilable patches. Patches should be sufficiently close to the buggy input program, but not so close that there is not significant change.

For each model, we exercise the ranking scheme on two sets of patches: ($a$) all generated patches, and ($b$)  the subset of generated patches that pass the compiler. 

To evaluate the ranking strategy, we choose the LLM configurations for each model with the highest discrepancy between pass@1 and pass@100 from RQ3 (see \Cref{tab:prompts}), which also happens to be the best performing configuration for each model.

\Cref{tab:ranking} shows the ranked pass@1 and ranked pass@5 accuracy (denoted by r.P@k) for the two sets of patches: 1) before (denoted as \texttt{All}) and 2) after (denoted as \texttt{Pruned}) pruning the compiler errors.
For reference, we also report the P@k metrics in parenthesis. 
We observe that pruning compiler errors does improve the metric P@k, especially the P@5 for \ic{davinci-002} model by over 14\% (33.97 up from 19.8).
Second, while providing determinism, r.P@1 improves over P@1 for most configurations except a slight dip for \gpt for \texttt{Pruned}. For \ic{davinci-002}, ranking improves the r.P@1 by over 5 percentage points.
Finally, r.P@5 remains close to P@5 for most configurations except for \ic{gpt-3.5} where the r.P@5 trails P@5 by 4.68\%. 

\RS{4}{By pruning compilation failures LLMs can achieve as high as 39.66\% pass@5. Using cosine similarity between embeddings of candidate patches and buggy input code, we can apply a deterministic ranking strategy that retains this high accuracy with 34.2\% - 35.68\% r.pass@5.}

\begin{table}[t!]
\caption{Accuracy of ranking with and without compiler pruning.}
\label{tab:ranking}
\centering
\resizebox{0.95\linewidth}{!}%
{
\begin{tabular}{@{}llccc@{}}
\toprule
\multirow{2}{*}{Approach} & \multirow{2}{*}{Metric}  & \ic{gpt-3.5} & \ic{davinci-edit} & \ic{davinci-002} \\
 &  & 1-shot & 0-shot & 0-shot \\ \midrule

\multirow{2}{*}{All} & r.P@1 (P@1) & 16.96 (16.22) & 13.78 (12.18) & 11.30 (6.29) \\
 & r.P@5 (P@5) & 31.8 (31.93) & 30.74 (28.86) & 17.66 (19.8) \\  \midrule

\multirow{2}{*}{Pruned}

& r.P@1 (P@1) & 21.20 (22.22) & 19.78 (18.99) & 17.66 (16.89)\\
 & r.P@5 (P@5) & 34.98 (39.66) & 35.68 (36.73) & 34.2 (33.97)\\ \bottomrule \\
\end{tabular}}
\end{table}

\section{Related Work}
\label{sec:related}

\newcommand{\nltofix}{{\it nl2Fix}}
\newcommand{\passAtk}{{\it pass@k}}

Our work is most closely related to two broad lines of work (a) automated code editing, and (b) automatic program repair (APR).

{\bf Automated Code Editing.} 
Earlier works in learning code editing include learning to edit code for refactoring~\cite{meng2015does, rolim2017learning}, learning semantic code changes for bugs found by code analyzers~\cite{rolim2018learning}. 
Recent approaches leverage Deep Learning techniques to learn frequent code edit patterns from code changes mined from GitHub~\cite{tufano2019learning, chakraborty2020codit, dinella2020hoppity}.

In addition to learning from historic changes, some recent approaches~\cite{chakraborty2021multi, zhang2022coditt5} propose to guide code editing with auxiliary inputs such as commit message. We argue that commit messages are {\em post-facto} description of the code changes, and do not capture the intent of the change, rather a summarization of the change.
In contrast, the issue report we consider in \nltofix{} problem is {\em ante-facto} description of the changes, which arguably capture the intent closer. 
Tufano~\etal~\cite{tufano2021towards} proposed automating code-review activity by editing the code based on reviewers comments. While closest to our work, the approaches differ in our usage of tests and semantic metrics such as pass@k to validate the functional correctness and defect-freedom of proposed patches.

{\bf Automatic Program Repair.}

Approaches for APR broadly fall under search-based techniques and machine learning-based methods.
For a comprehensive overview of automated program repair techniques, we refer readers to recent works that survey the area \cite{apr_survey_monperrus_acm_2018,apr_survey_gazzola_2019,goues2019automated,gao2022program}.

Search-based techniques use a {\it generate-and-validate} approach, where variations of the original code are generated and then evaluated using the failing tests\cite{saha2017elixir, weimer2013leveraging, qi2014strength, tan2015relifix, kim2013automatic}. These approaches transform buggy code using different transformation including random transformation~\cite{weimer2013leveraging, genprog, qi2014strength}, manually designed transformation~\cite{saha2017elixir}, and transformation learned from previous corpus of code~\cite{kim2013automatic, le2016history,tufano2018empirical, chen2019sequencer}. 

Recently, researchers have also leveraged LLMs for APR.
Alpharepair~\cite{alpharepair-22} uses a {\it cloze}-style APR where an LLM directly fills-in the correct code given its surrounding context from the buggy program.
Xia et al.~\cite{xia2023conversational} uses ChatGPT to setup a conversational APR problem where feedback from failed patches is used to augment subsequent prompts to LLMs. 
Finally, there are approaches leveraging LLMs for generating trigger tests for APR problem from Issue descriptions~\cite{kang2022large} as well as aiding rootcausing for APR~\cite{Motwani23icse}. Finally, Fan et al.~\cite{fan2023automated} leverage APR methods (including those based on LLMs) to repair code generated from natural language intent.

Although closely related, our approach subtly differs from  APR in the use of {\it hidden tests} that are only used for evaluation and never specified as inputs to the repair algorithm. 
This makes it a more applicable problem for real-world bug fixes that may not have failing tests available (or be prohibitively expensive to leverage) during the inference time for help in rootcausing buggy lines and validating the fixes.

\section{Limitations and Threats}
\label{sec:threats}

\emph{Stability of models' output.} As we have used OpenAI web API to access different models, we cannot control the stochasticity of the output by the model. The models themselves are often updated. This poses a threat to the replicability of our study. To mitigate this threat, we make available all the outputs generated by the models. 

\emph{Assumption on patch correctness.} 
In this paper, we leverage tests to determine if a fix is plausible. However, such a plausible patch may not fix a bug completely. This may need manual analysis or a semantic equivalence with the user-provided fix. However, the former is infeasible for a large-scale benchmark such as \dforjnltoedit{}, and semantic equivalence-checking techniques do not scale to handle most real programs. 
Given that test-suites are never exhaustive, we can appeal to recent research that investigates patch-correctness~\cite{le2019reliability, xiong2018identifying, wang2020automated} to improve confidence on the patches.

\emph{Generalization of findings.}
Given the relatively small number of bugs (283) considered in \dforjnltoedit{} benchmark, our findings may not generalize to arbitrary bugs across different languages and software repositories. To mitigate this threat we use real-world bugs from open source projects.

\section{Conclusion}
\label{sec:conclusion}

In this paper, we motivate the \nlTofix{} problem, define the first benchmark \dforjnltoedit{}, and perform detailed empirical evaluation of various SOTA LLMs on the problem.

We believe that the task of \nlTofix{} along with challenging benchmarks such \dforjnltoedit{} will serve as an important real-world benchmark for evaluating future generation of these LLMs (such as GPT-4), while leveraging new emergent behaviors of such LLMs (such as the ability of these models to predict the correctness) to improve the performance on such benchmarks. 
In future work, we plan to combine issue-driven trigger test generation~\cite{kang2022large} and user-in-the-loop~\cite{ticoder} to improve the trust in generated fixes, as well as extend our framework to the more general problem of \nlToedit{}  to cover other forms of program evolution including feature additions, refactorings as well as optimizations.

\bibliography{main}

\end{document}